\documentclass[sn-mathphys-num]{sn-jnl}

\usepackage{graphicx}%
\usepackage{multirow}%
\usepackage{amsmath,amssymb,amsfonts}%
\usepackage{amsthm}%
\usepackage{mathrsfs}%
\usepackage[title]{appendix}%
\usepackage{xcolor}%
\usepackage{textcomp}%
\usepackage{manyfoot}%
\usepackage{booktabs}%
\usepackage{algorithm}%
\usepackage{algorithmicx}%
\usepackage{algpseudocode}%
\usepackage{listings}%
\usepackage[utf8]{inputenc}  
\usepackage[T1]{fontenc}     
\usepackage{lmodern}         

\begin{document}

\title[Outcomes from a Workshop on a National Center for Quantum Education]{Outcomes from a Workshop on a National Center for Quantum Education}


\author[1]{\fnm{Edwin} \sur{Barnes}}\email{efbarnes@vt.edu}
\affil[1]{\orgdiv{Department of Physics and Virginia Tech Center for Quantum Information Science and Engineering}, \orgname{Virginia Tech}, \orgaddress{\street{850 West Campus Drive}, \city{Blacksburg}, \postcode{24061}, \state{VA}, \country{USA}}}

\author[2]{\fnm{Michael B.} \sur{Bennett}}\email{michael.bennett@colorado.edu}
\affil[2]{\orgdiv{Q-SEnSE NSF QLCI}, \orgname{University of Colorado}, \orgaddress{\street{440 UCB}, \city{Boulder}, \postcode{80301}, \state{CO}, \country{USA}}}

\author[3,4]{\fnm{Alexandra} \sur{Boltasseva}}\email{aeb@purdue.edu}
\affil[3]{\orgdiv{Elmore Family School of Electrical and Computer Engineering, Birck Nanotechnology Center and Purdue Quantum Science and Engineering Institute}, \orgname{Purdue University}, \orgaddress{\street{610 Purdue Mall}, \city{West Lafayette}, \postcode{47907}, \state{IN}, \country{USA}}}
\affil[4]{\orgdiv{Quantum Science Center, a National Quantum Information Science Research Center of the U.S. Department of Energy}, \orgname{Oak Ridge National Laboratory}, \orgaddress{\street{1 Bethel Valley Road}, \city{Oak Ridge}, \postcode{37830}, \state{TN}, \country{USA}}}

\author[5,6]{\fnm{Victoria} \sur{Borish}}\email{victoria.borish@colorado.edu}
\affil[5]{\orgname{JILA, National Institute of Standards and Technology and the University of Colorado}, \orgaddress{\city{Boulder}, \state{Colorado}, \postcode{80309},  \country{USA}}}
\affil[6]{\orgdiv{Department of Physics}, \orgname{University of Colorado}, \orgaddress{\street{390 UCB}, \city{Boulder}, \state{Colorado}, \postcode{80309}, \country{USA}}}

\author[7]{\fnm{Bennett} \sur{Brown}}\email{bbrown@qusteam.org}
\affil[7]{\orgname{QuSTEAM}, \orgaddress{\street{510 Ronalds St.}, \city{Iowa City}, \postcode{52245}, \state{IA}, \country{USA}}}

\author[8]{\fnm{Lincoln D.} \sur{Carr}}\email{lcarr@mines.edu}
\affil[8]{\orgdiv{Quantum Engineering Program and Department of Physics}, \orgname{Colorado School of Mines}, \orgaddress{\street{1523 Illinois St.}, \city{Golden}, \postcode{80401}, \state{CO}, \country{USA}}}

\author[9]{\fnm{Russell R.} \sur{Ceballos}}\email{rceballos@qusteam.org}
\affil[9]{\orgdiv{QuSTEAM, non-profit}, \orgname{Wilbur Wright College - City Colleges of Chicago}, \orgaddress{\street{4300 N Narragansett Ave}, \city{Chicago}, \postcode{60634}, \state{IL}, \country{USA}}}

\author[10]{\fnm{Faith} \sur{Dukes}}\email{fmdukes@lbl.gov}
\affil[10]{\orgdiv{Government and Community Relations Office - K-12 Programs}, \orgname{Lawrence Berkeley National Laboratory}, \orgaddress{\street{}, \city{1 Cyclotron Road}, \postcode{Berkeley}, \state{CA}, \country{USA}}}

\author[11]{\fnm{Emily W.} \sur{Easton}}\email{eweaston@uchicago.edu}
\affil[11]{\orgname{Chicago Quantum Exchange}, \orgaddress{\street{5235 S. Harper Court}, \city{Chicago}, \postcode{60615}, \state{IL}, \country{USA}}}

\author[12]{\fnm{Sophia E.} \sur{Economou}}\email{economou@vt.edu}
\affil[12]{\orgdiv{Department of Physics}, \orgname{Virginia Tech}, \orgaddress{\street{850 West Campus Drive}, \city{Blacksburg}, \postcode{24060}, \state{VA}, \country{USA}}}

\author[13]{\fnm{E. E.} \sur{Edwards}}\email{emily.edwards@duke.edu}
\affil[13]{\orgdiv{Department of Electrical and Computer Engineering and Duke Quantum Center, Duke University}, \orgname{Duke University}, \orgaddress{\street{130 Hudson Hall}, \city{Durham}, \postcode{27708}, \state{NC}, \country{USA}}}

\author[14]{\fnm{Noah D.} \sur{Finkelstein}}\email{noah.finkelstein@colorado.edu}
\affil[14]{\orgdiv{Department of Physics}, \orgname{University of Colorado}, \orgaddress{\street{390 UCB}, \city{Boulder}, \state{Colorado}, \postcode{80309}, \country{USA}}}

\author[15]{\fnm{C.} \sur{Fracchiolla}}\email{fracchiolla@aps.org}
\affil[15]{\orgname{American Physical Society}, \orgaddress{\street{555 12th St. NW}, \city{Washington}, \postcode{20004}, \state{DC}, \country{USA}}}

\author[16]{\fnm{Diana} \sur{Franklin}}\email{dmfranklin@uchicago.edu}
\affil[16]{\orgdiv{Computer Science}, \orgname{University of Chicago}, \orgaddress{\street{5730 S. Ellis Ave.}, \city{Chicago}, \postcode{60637}, \state{IL}, \country{USA}}}

\author[17]{\fnm{J. K.} \sur{Freericks}}\email{james.freericks@georgetown.edu}
\affil[17]{\orgdiv{Department of Physics}, \orgname{Georgetown University}, \orgaddress{\street{552 Reiss Science Building}, \city{Washington}, \postcode{20057}, \state{DC}, \country{USA}}}

\author[18]{\fnm{Valerie} \sur{Goss}}\email{vgoss@csu.edu}
\affil[18]{\orgdiv{Chemistry, Physics and Engineering Studies}, \orgname{Chicago State University}, \orgaddress{\street{9501 S. King Drive, WSC 309}, \city{Chicago}, \postcode{60628}, \state{IL}, \country{USA}}}

\author[19]{\fnm{Mark} \sur{Hannum}}\email{mshannum@fcps.edu}
\affil[19]{\orgdiv{Physics}, \orgname{Thomas Jefferson High School for Science and Technology}, \orgaddress{\street{6560 Braddock Road}, \city{Alexandria}, \postcode{22312}, \state{VA}, \country{USA}}}

\author[20]{\fnm{Nancy} \sur{Holincheck}}\email{mshannum@fcps.edu}
\affil[20]{\orgdiv{College of Education and Human Development}, \orgname{George Mason University}, \orgaddress{\street{4400 University Drive}, \city{Fairfax}, \postcode{22030}, \state{VA}, \country{USA}}}

\author[21]{\fnm{Angela M.} \sur{Kelly}}\email{angela.kelly@stonybrook.edu}
\affil[21]{\orgdiv{Department of Physics and Astronomy}, \orgname{Stony Brook University}, \orgaddress{\street{100 Nicolls Road}, \city{Stony Brook}, \postcode{11794-5233}, \state{NY}, \country{USA}}}

\author[22]{\fnm{Olivia} \sur{Lanes}}\email{olivia.lanes@ibm.com}
\affil[22]{\orgname{IBM Quantum}, \orgaddress{\street{1101 Kitchawan Rd}, \city{Yorktown Heights}, \postcode{10598}, \state{NY}, \country{USA}}}

\author*[23,24]{\fnm{H.~J.} \sur{Lewandowski}}\email{lewandoh@colorado.edu}
\affil*[23]{\orgname{JILA, National Institute of Standards and Technology and the University of Colorado}, \orgaddress{\city{Boulder}, \state{Colorado}, \postcode{80309},  \country{USA}}}
\affil[24]{\orgdiv{Department of Physics}, \orgname{University of Colorado}, \orgaddress{\street{390 UCB}, \city{Boulder}, \state{Colorado}, \postcode{80309}, \country{USA}}}

\author[25]{\fnm{Karen Jo} \sur{Matsler}}\email{kmatsler@uta.edu}
\affil[25]{\orgdiv{College of Science, UTeach}, \orgname{University of Texas Arlington}, \orgaddress{\street{502 Yates St/SH 224}, \city{Arlington}, \postcode{76019}, \state{TX}, \country{USA}}}

\author[26]{\fnm{Emily} \sur{Mercurio}}\email{mercurio@umd.edu}
\affil[26]{\orgdiv{Quantum Leap Challenge Institute for Robust Quantum Simulation}, \orgname{University of Maryland}, \orgaddress{\street{Stadium Drive}, \city{College Park}, \postcode{20742}, \state{MD}, \country{USA}}}

\author[27,28]{\fnm{Inès} \sur{Montaño}}\email{ines.montano@nau.edu}
\affil[27]{\orgdiv{Department of Applied Physics and Materials Science}, \orgname{Northern Arizona University}, \orgaddress{\street{527 S Beaver St}, \city{Flagstaff}, \postcode{86011}, \state{AZ}, \country{USA}}}
\affil[28]{\orgdiv{Center for Materials Interfaces in Research and Applications}, \orgname{Northern Arizona University}, \orgaddress{\street{527 S Beaver St}, \city{Flagstaff}, \postcode{86011}, \state{AZ}, \country{USA}}}

\author[29]{\fnm{Maajida} \sur{Murdock}}\email{maajida.murdock@morgan.edu}
\affil[29]{\orgdiv{Department of Physics and Engineering Physics}, \orgname{Morgan State University}, \orgaddress{\street{1700 East Cold Spring Lane, Calloway Hall - G22}, \city{Baltimore}, \postcode{21251}, \state{MD}, \country{USA}}}

\author[30]{\fnm{Kiera} \sur{Peltz}}\email{kiera@the-cs.org}
\affil[30]{\orgname{The Coding School}, \orgaddress{\street{6940 Laurel Canyon Blvd}, \city{Los Angeles}, \postcode{91604}, \state{CA}, \country{USA}}}

\author[31]{\fnm{Justin K.} \sur{Perron}}\email{jperron@csusm.edu}
\affil[31]{\orgdiv{Department of Physics}, \orgname{California State University San Marcos}, \orgaddress{\street{333 S. Twin Oaks Valley Rd}, \city{San Marcos}, \postcode{92069}, \state{CA}, \country{USA}}}

\author[32]{\fnm{Christopher J.K.} \sur{Richardson}}\email{richardson@lps.umd.edu}
\affil[32]{\orgname{Laboratory for Physical Sciences}, \orgaddress{\street{8050 Greenmead Dr.}, \city{College Park}, \postcode{20740}, \state{MD}, \country{USA}}}

\author[33]{\fnm{Jessica L.} \sur{Rosenberg}}\email{jrosenb4@gmu.edu}
\affil[33]{\orgdiv{Department of Physics and Astronomy}, \orgname{George Mason University}, \orgaddress{\street{4400 University Drive}, \city{Fairfax}, \postcode{22030}, \state{VA}, \country{USA}}}

\author[34]{\fnm{Richard S.} \sur{Ross}}\email{rsross@physics.ucla.edu}
\affil[34]{\orgdiv{Department of Physics and Astronomy}, \orgname{University of California Los Angeles}, \orgaddress{\street{475 Portola Plaza}, \city{Los Angeles}, \postcode{90095}, \state{CA}, \country{USA}}}

\author[35]{\fnm{Minjung} \sur{Ryu}}\email{mjryu@uic.edu}
\affil[35]{\orgdiv{Chemistry, Learning Sciences Research Institute}, \orgname{University of Illinois Chicago}, \orgaddress{\street{845 W Taylor St.}, \city{Chicago}, \postcode{60607}, \state{IL}, \country{USA}}}

\author[36]{\fnm{Raymond E.} \sur{Samuel}}\email{resamuel@ncat.edu}
\affil[36]{\orgdiv{Department of Biology}, \orgname{North Carolina A and T State University}, \orgaddress{\street{1601 E Market St.}, \city{Greensboro}, \postcode{27411}, \state{NC}, \country{USA}}}

\author[37]{\fnm{Nicole} \sur{Schrode}}\email{schrode@aps.org}
\affil[37]{\orgname{American Physical Society}, \orgaddress{\street{One Physics Ellipse}, \city{College Park}, \postcode{20740}, \state{MD}, \country{USA}}}

\author[38]{\fnm{Susan} \sur{Schwamberger}}\email{susan.schwamberger@quantinuum.com}
\affil[38]{\orgname{Quantinuum}, \orgaddress{\street{03 S Technology Court,}, \city{Broomfield}, \postcode{80021}, \state{CO}, \country{USA}}}

\author[39]{\fnm{Thomas A.} \sur{Searles}}\email{tsearles@uic.edu}
\affil[39]{\orgdiv{Department of Electrical and Computer Engineering}, \orgname{University of Illinois Chicago}, \orgaddress{\street{851 S. Morgan St.}, \city{Chicago}, \postcode{60607}, \state{IL}, \country{USA}}}

\author[40]{\fnm{Chandralekha } \sur{Singh}}\email{clsingh@pitt.edu}
\affil[40]{\orgdiv{Department of Physics}, \orgname{University of Pittsburgh}, \orgaddress{\street{100 Allen hall 3941 Ohara St.}, \city{Pittsburgh}, \postcode{15260}, \state{PA}, \country{USA}}}

\author[41]{\fnm{Alexandra} \sur{Tingle}}\email{alexandra.tingle@gmail.com}
\affil[41]{\orgname{Infleqtion}, \orgaddress{\street{1315 W Century Dr. 150}, \city{Louisville}, \postcode{80027}, \state{CO}, \country{USA}}}

\author[42]{\fnm{Benjamin M.} \sur{Zwickl}}\email{ben.zwickl@rit.edu}
\affil[42]{\orgdiv{School of Physics and Astronomy}, \orgname{Rochester Institute of Technology}, \orgaddress{\street{84 Lomb Memorial Dr.}, \city{Rochester}, \postcode{14623}, \state{NY}, \country{USA}}}

\abstract{In response to numerous programs seeking to advance quantum education and workforce development in the United States, experts from academia, industry, government, and professional societies convened for a National Science Foundation-sponsored workshop in February 2024 to explore the benefits and challenges of establishing a national center for quantum education. Broadly, such a center would foster collaboration and build the infrastructure required to develop a diverse and quantum-ready workforce. The workshop discussions centered around how a center could uniquely address gaps in public, K-12, and undergraduate quantum information science and engineering (QISE) education. Specifically, the community identified activities that, through a center, could lead to an increase in student awareness of quantum careers, boost the number of educators trained in quantum-related subjects, strengthen pathways into quantum careers, enhance the understanding of the U.S. quantum workforce, and elevate public engagement with QISE. Core proposed activities for the center include professional development for educators, coordinated curriculum development and curation, expanded access to educational laboratory equipment, robust evaluation and assessment practices, network building, and enhanced public engagement with quantum science. 
}
\keywords{quantum, education, workforce}

\maketitle

\section{Introduction}\label{sec:intro}

Historically, formal “quantum education” has been available via conventional undergraduate and graduate coursework in quantum mechanics, physical chemistry, optics, and other related topics in modern physics and electrical engineering. Recent analyses of the landscape support this assertion, and also capture data on coursework relating to quantum information science and engineering (QISE) \cite{Wilcox2024}. For instance, the number of quantum-related courses varies across institutions and thus not all undergraduates have access to the same options. In recent years, a handful of online offerings and informal resources complement the more formal coursework, opening up more entry points into the field (see \cite{Q12}, \cite{qureca2022}, for example). Spurred by global investments in QISE, experts from academia, government, and industry have convened multiple meetings, and identified critical challenges related to the educational landscape, which in turn drives the current and future quantum workforce. In the United States (US), such challenges include, but are not limited to, the lack of curricula at the K-12 and lower-division undergraduate levels, insufficient educator readiness and availability, variable support for educator professional development, and inconsistent course offerings.

Notably, the current sequence in which quantum concepts are offered in formal learning is not necessarily designed to provide equitable or even expanded access to quantum careers or nascent quantum technologies. Additionally, the current model does not support broadening public awareness of QISE, as most students never encounter it directly in formal classrooms. Therefore, if the goal is to effectively support a larger, diverse domestic workforce with varying levels of quantum proficiency, then it is necessary to focus on expanding quantum education offerings at the K-12 and undergraduate levels.

During previous meetings, members of the community have proposed interventions such as pre-college curricula on quantum-related phenomena, new undergraduate minors, and professional certifications \cite{Asfaw2022}. Additionally, various groups are focused on addressing the lack of quantum opportunities for historically marginalized students through, for instance, the establishment of a center that connects quantum industry directly to Historically Black Colleges and Universities (HBCUs) \cite{lee2021}, federally-funded opportunities (e.g., NSF ExpandQISE and DOE RENEW), and by exploring methodologies for teaching QISE to students earlier and in more settings \cite{games-OE, qcamp}. The lack of expansive up-to-date data on the quantum education and workforce landscape, coupled with the long timeline for making changes, and tracking their effectiveness, complicates sustaining nascent programs that show early signs of success.

To ensure continued progress, the emergent quantum educational ecosystem in the US has, in recent years, rallied around the need for a center-scale investment focused on cohering the field of quantum education and accelerating implementation. In this context, a center refers to an investment of greater than \$5M per year and incorporating experts from multiple institutions, and may be called a hub or other term. For simplicity, this paper refers to this type of investment as a center. Government-funded research and education centers tend to bring together a set of experts to tackle a challenge under the following general circumstances: (1) a discipline is new, and thus there is limited capacity among experts, (2) there is a need for long-term solutions that a single entity cannot provide, and (3) there is a challenge spanning multiple sectors, yet no commercial or sustained market-driven solution is available. As this paper will discuss, quantum education prior to graduate school meets all of these criteria. The community call-to-action for sustained, significant investment is aligned with US national strategy documents; the Quantum Information Science and Technology Workforce Development National Strategic Plan calls for collaboration amongst federal agencies, K-12 schools, colleges, universities, and informal institutions to support an increasing and diverse quantum educational opportunities. \cite{Quantum}. Specifically, the "Four critical actions'' listed in the strategy must be implemented at the K-12 and undergraduate levels for US to be effective at expanding the quantum-ready workforce. 

To better understand the community call for a large-scale investment, faculty from the University of Colorado Boulder and Duke University organized a workshop in early 2024 to discuss and document the opportunities and need for a center-scale investment in quantum education within the context of the US research and educational ecosystem. This paper reports on the outcomes of that workshop, which was supported by the National Science Foundation (NSF). Attendees included experts from K-12 education (pre-college), higher education (including community colleges and four-year institutions), industry (for profit and non-profit), government, and professional societies. Specifically, attendees met to (1) outline possible activities that a center-scale approach could enable, (2) discuss challenges, (3) propose center structures that align with existing STEM (Science, Technology, Engineering, and Mathematics) education investments, and (4) outline possible complementary existing and future investments that could enhance the center and ecosystem. In the geographical and cultural context of the US, scaled and sustained approaches to quantum education spanning K-12 through undergraduate (K-16) are especially necessary to tap into the pool of diverse learners who have historically been left out of technology revolutions of the last century \cite{Asfaw-PT, Lee-2023, perron-2021}. Here and throughout the paper, K-12 refers to students typically ages 5-18 and undergraduate is inclusive of 2- and 4-year colleges. In the US, 2-year colleges are often dubbed `community colleges' with coursework that may be tailored to local employment needs, and students can receive technical certificates, associates degrees, and transfer to 4-year programs. The National Center on Education Statistics provides additional information on the various classifications of undergraduate programs \cite {ipeds}.

While the workshop and findings are tailored to the US, this paper provides a framework of activities that could be organized and adapted to other global contexts, recognizing that there are significant differences in education and governance systems worldwide. Similarly, many industries seeking quantum-conversant employees are multi-national and share a common interest in supporting the development of robust quantum education. Therefore, the findings may facilitate the strategic participation of companies in educational program development and implementation.

\section{Current Related Workforce Development Activities}\label{sec:background}

We distinguish between quantum education and workforce development in order to provide clarity on the workshop findings. Here, quantum education refers to the formal and informal learning opportunities that students spanning K-16 through graduate school could potentially access. In this context, quantum education is one aspect of a broader strategy of workforce development that goes beyond training students for specific roles within a single sector.

Currently, most of the US's  activity in quantum education operates via five main mechanisms: National QISE research centers, small-team or single investigator academic programs, private-sector courses and tutorials, non-profits, and conventional STEM departments. The national research centers were designed to focus on activities that accelerate real-world quantum applications. In doing so, these centers prepare graduate students and limited numbers of undergraduates for their future careers through research, summer schools, professional skills-building (e.g., career fairs). Since 2020, governmental agencies have made concerted efforts to also support small teams to research and design educator professional development, experiential learning, curricula, and other educational deliverables. Private-sector companies have also developed a mix of free and paid online tutorials on quantum computing, quantum-safe cryptography, and quantum algorithms, for example. While each of these partially fills a gap in the educational ecosystem, the current and anticipated labor trends necessitate new and sustained educational programs that ultimately increase the number of people with knowledge and skills in QISE from an earlier stage \cite{Hughes2022,Fox2020,Aiello_2021}. 

Educators, employers, and researchers have also recognized a need to better connect and coordinate quantum education efforts to understand their impact on developing a workforce. Starting in 2019, there have been several meetings related to broadly developing quantum education and workforce development programs that laid the groundwork for the workshop described in this paper. 
\begin{itemize}
    
    \item Kavli Futures Symposium: Achieving a Quantum Smart Workforce Nov 4-5 2019 \cite{Aiello_2021}
    \item Key Concepts for Future QIS Learners, March-April 2020 \cite{KC}
    \item QUEST: Quantum Undergraduate Education and Scientific Training, Jun 3 2021 \cite{perron2021quest}
    \item Q-12 Actions For Community Growth, Feb 1 2022 \cite{q12-meet}
    \item NSF Workshop on Quantum Engineering Education, Feb 25-26 2021 \cite{Asfaw2022}
    \item Workshop on Quantum Education for Quantum Workforce Development, Jan 29-31 2023
    \item Supporting Minority Serving Institutions in the Creation of a Diverse, Quantum-Ready Workforce, July 13 2023
\end{itemize}

In addition, IEEE holds a quantum education conference during its annual ''Quantum Week" and the APS March Meeting regularly features multiple sessions on quantum education. Lastly, the US National Quantum Coordination Office developed a strategy on the quantum workforce \cite{NQCC2022}, the release of which coincided with item four above. One common theme from these conferences, workshops, papers, and reports is the call to prioritize changes to pre-college and undergraduate education such that QISE is infused into existing coursework, and course sequences are modified as appropriate. Acting on these calls will require a long-term investment in quantum education. The \textit{Workshop on a National Center for Quantum Education and Workforce Development} that this paper reports on represents the first time that experts across the ecosystem have gathered to conceptualize a dedicated, center-scale, national resource focused on addressing challenges in quantum education at scale, over a long period of time.

\section{Gathering of Community Input}\label{sec:workshop}

The main goal of the workshop was to gather input from a wide range of experts, including educators, industry professionals, and representatives from professional societies and government agencies, on the needs, challenges, and opportunities of a center-scale investment in quantum education in the US. 

The organizing committee, consisting of Emily Edwards, Noah Finkelstein, and Heather Lewandowski, developed an invitation list with input from leaders across the community to ensure broad representation. To facilitate significant interactions among attendees, the workshop was by invitation only and limited to around 40 participants. Invitees included national leaders in quantum education efforts, educators working on small-team quantum education projects, participants of previous workshops, representatives from quantum companies, and relevant professional societies. The final invite list was carefully curated to ensure diversity in sector representation, geography, and demographics. A total of 53 people were invited, and 35 attended the in person portion. The attendees represented a broad range of disciplines including physics, engineering, chemistry, and computer science and areas of expertise including public engagement, education research, informal science education, and formal education across all levels. There were also representatives from companies ranging from small start-ups to large established industry leaders and several national labs. 

Since several people could not attend the in-person workshop, the organizing committee followed up with three 2-hour Zoom sessions to accommodate additional participants' schedules. Around 20 individuals attended one of these virtual sessions. During these meetings, the organizing committee presented the workshop's framing, preliminary outcomes, and key discussion points. Participants provided input on these outcomes, including additional components that might have been missed during the workshop.

The 1.5-day in-person workshop was held at JILA at the University of Colorado Boulder on February 12 --13, 2024. Based on the previous meetings, the National Quantum Initiative Advisory Committee (NQIAC) report from June 2, 2023 \cite{NQiac2023}, and information from the June 7, 2022 National Quantum Initiative Reauthorization Act (NQIRA) public hearing of the US House of Representatives Committee on Science, Space, and Technology, the workshop organizers developed the agenda shown below. 

 The workshop included five presentations to establish the background on the current state of quantum education across the US and framing of the workshop goals. The slides from these presentations were shared with the virtual participants.
\begin{itemize}
    \item Workshop on a national quantum education center
    \item Overview of early quantum education landscape
    \item Overview of undergraduate quantum education landscape
    \item Overview of different types of investments for supporting scaling
    \item Challenges and Success of Partnerships Beyond Academic Enterprise
\end{itemize}

Aside from these presentations, in-person participants spent most of the first day in  small discussion groups followed by full group discussions organized into three main sessions that centered around the following questions:

\begin{itemize}
    \item What activities most benefit from a center-scale investment in quantum education and workforce development at different educational institutions? What challenge does each activity address?
    \item What models of investment will best support the activities/challenges at scale? 
    \item How else can industry, government labs, and other employers partner and actively contribute to address challenges/activities? 
\end{itemize}

The final half day was dedicated to organizing the points of agreement, and consensus when possible, and outcomes from the first day of the workshop.

\section{Findings}\label{sec:findings}

\subsection{Need, Scope, and Structure for a Center}

During the workshop, attendees discussed the necessity of a center-scale investment in the context of the current landscape, assuming that small projects would continue. Attendees widely agreed that there remains a need for significant, new national infrastructure to support quantum education as part of possible future reauthorizations of the National Quantum Initiative Act (NQIA) or other related legislation. In short, attendees felt this type of investment is needed for the same reason that there are centers focused on scientific research -- to enable deeper collaboration on a critical challenge, and to drive innovation and new approaches. During small group discussions, workshop participants outlined the essential elements, approaches, and scale for an education-focused center or hub. The activities and outcomes are listed below in subsequent sections. 

Attendees also discussed the complexity of workforce development in a nascent area, and how to scope a center that would be impactful, but not necessarily comprehensive. Crucially, the attendees agreed that a center should focus on primarily the K-12 and undergraduate levels, while including clear connections to graduate education, technical training, and employers. The sections on activities and outcomes that follow reflect this proposed focus. Including both K-12 and undergraduate education in a single center recognizes that (1) the success of students in STEM courses at the undergraduate level depends partially on their preparation from high school; (2) from a planning standpoint there is an interplay between curriculum in K-12 and introductory undergraduate courses; (3) quantum curricula can be adapted for different courses and ages; and (4) educators at all levels benefit from the connections spanning K-16. Some participants were concerned about an education center being to broad or sprawling, which would limit its effectiveness. To partially address this, most agreed that a center focused on K-16 should limit its activities and programming in other areas of education and workforce development (e.g., master's programs, technical reskilling) in order to have a manageable and effective scope of work at the earlier levels. Even so, some participants highlighted that such programs will likely continue to evolve and support additional pathways into quantum-related careers. Thus, while most cohered around a focused K-16 center, this conclusion should not preclude future investments in these other domains.

Structurally, the workshop attendees envisioned a center as both geographically and institutionally distributed, designed to support the range of activities and associated outcomes described later in this paper. A distributed center could, for instance, include regional clusters that link K-12, community colleges and universities, local and national industry, and the public (e.g., through community partners). Connections between clusters could form across common themes (e.g., professional development and curricula). Ultimately, a national infrastructure, scaffolded by local expertise, could accelerate collaborative solutions that subsequently could be adapted to different geographic and institutional needs. Additionally, a center would be well-positioned to analyze and disseminate changing employer needs, as well as monitor national trends (e.g., demographics, population distribution).

 \subsection{Possible Center Activities}
Attendees considered many activities that could fall within a K-16 center and would benefit from this type of scaled collaboration. They recommended focusing on areas that would broadly increase the overall \textit{quality of} and \textit{capacity for} QISE education, and fill persistent gaps in the ecosystem. They suggested a strong emphasis on supporting professional development (PD) for educators, curriculum development, public engagement, and analysis of the educational and workforce landscape. Each of these activities would be approached in a scholarly manner, drawing from what is known and contributing new knowledge through evaluation and research.  While these efforts are categorized below as separate activities, workshop attendees envisioned an integrated approach. For example, integrating curricular development, PD, and education research will ensure that outputs are co-designed with educators and evidenced-based instructional strategies tested for efficacy. Following this section, we outline the collective outcomes that a comprehensive approach to a center could accomplish. This is summarized in Fig 1.

\begin{figure}
    \centering
    \includegraphics[width=1\linewidth]{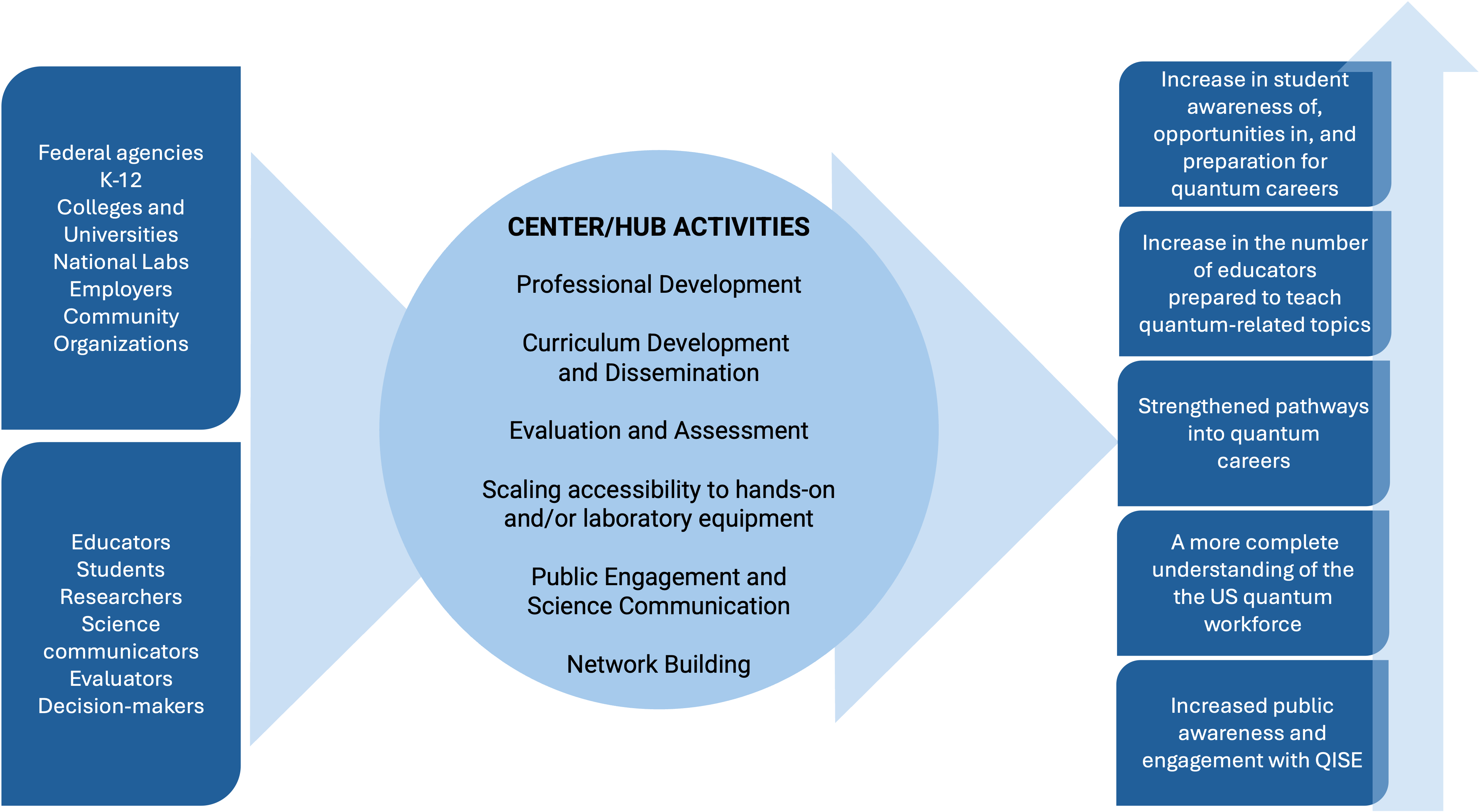}
    \caption{Diagram illustrating the key activities of a proposed center or hub for Quantum Information Science and Engineering (QISE) and their expected outcomes. The center's activities could include professional development, curriculum development and dissemination, evaluation and assessment, scaling access to hands-on and laboratory equipment, public engagement, science communication, and network building. The engagement involves diverse stakeholders such as federal agencies, K-12 institutions, universities, national labs, employers, educators, researchers, and community organizations. These activities aim to increase (indicated by the arrow) student awareness of quantum careers, boost the number of educators trained in quantum-related subjects, strengthen pathways into quantum careers, enhance the understanding of the U.S. quantum workforce, and elevate public engagement with QISE.}
    \label{fig:enter-label}
\end{figure}

\textbf{ Professional Development (PD}) A majority of participants felt that a center should incorporate a strong focus on preparing educators at all levels (K-16) to develop both the necessary content knowledge and associated pedagogical expertise. In addition to providing PD, by including K-12 and undergraduate educators, a center could better facilitate educator awareness of the range of future educational opportunities available to their students.

There was clear consensus that individual or small programs most likely cannot deploy PD either at scale or for multiple years without the sustained scaffolding a center could provide. For example, in the US, many new programs are funded for three years and tied to a particular geographic location. Typically, this initial investment is sufficient to develop initial institutional (e.g., school district, university) support and engage in educational development with small to medium cohorts with no more than 10--100 educators. However, there is difficulty in scaling beyond the initial funding period. A center could begin to facilitate a solution to this issue by sharing research-based practices across a wider community of participants, and providing connectivity among PD programs. This could potentially increase the pool of educators and lessen the burden of 'starting from scratch' with each new individual program. In areas where there are gaps, a center could seed new activity. For example, some programs have experienced challenges in recruiting computer science educators for PD for quantum education. A center could help address this by onboarding additional partner organizations in computer science education who could potentially test out lesson modules and co-organize PD workshops in QISE. 
     
 Participants acknowledge that in practice PD will look different for K-12 versus undergraduate educators (both 2-year and 4-year), but that each group would benefit from materials and practices developed across educational levels. For example, there is currently a new program in California that supports the development of QISE faculty learning communities within primarily undergraduate education \cite{perron2021quest}. While the curricula may vary, the model they use for supporting educators online throughout the year could benefit the K-12 quantum education community. Participants also noted that building community as a component of PD is especially important for supporting educators learning to teach a new discipline, even if the activities are virtual and therefore more cost-effective.  

In all cases, workshop attendees noted the need for incentives, including financial and travel support, for educators participating in professional development. In K-12 specifically, compensated PD is standard practice and can cost anywhere between a few hundred dollars and several thousand per participant, depending on the length of PD and type (workshop versus research experience). In many cases, PD also counts towards required credits in continuing education, providing additional incentive. For undergraduate educators, it may be more common to provide travel support to conferences or to other institutions for visiting programs. It is likely infeasible that a single center could provide compensation for tens of thousands of educators across the country but, a center could sustain cohorts across a longer time period, provide additional seed funding in strategic areas, and create critical visibility and connectivity necessary for attracting further investments and partnerships from the larger STEM ecosystem. At the K-12 level, if topics are eventually incorporated into pre-service education and integrated into post-service STEM PD, then the requirement for PD, driven by a center as described here, may be reduced. 

\textbf{Collaboration on Curriculum and Resource Development and Dissemination} Currently, there is increasing activity across all levels in curriculum and activity development in formal and informal learning environments, respectively. Curricula in this context refers to learning materials, information on how to use them, assessments, learning outcomes, and other formal scaffolding. Curricula can be offered in formal courses or informal environments such as online (e.g., edX).  However, practitioners may have limited connectivity, and there are barriers to access many of the resources (e.g., paywalls, membership requirements). Attendees articulated that implementation appears to be uneven and located at few institutions or programs. A center could enable collaboration, help avoid duplicate efforts, and provide a way to leverage knowledge across the education community to establish effective practices for teaching QISE across more institutions, educational levels, and learning environments. 

Attendees discussed the tension between centralized curriculum development and independence for instructors at different levels. For example, professors in higher education often develop their own lectures and will not necessarily utilize a fully pre-packaged curriculum. K-12 educators may seek to incorporate only a few lessons, rather than a full-scale quantum course. Yet both groups would benefit from a database of vetted resources (e.g., courses or modules) that span the breadth of the field and have options for tie-ins to different types of science, engineering, and technology coursework. 

A center could also periodically analyze the gaps in curricula and support new resources as appropriate (e.g., a seed program). For example, a center could enable collaboration on a set of comprehensive learning outcomes and expectations for different areas of QISE, and articulate overlaps and distinctions, building on prior work (see \cite{KC} and the QIS Framework \cite{Q12}, for example). This could lay a foundation for learning modules across the full spectrum of the field, filling in gaps where there may be little to no curricula at present. Attendees agreed that a center should complement and develop synergies with the efforts of smaller teams, individual grantees, non-profits, and quantum companies, and not seek to centralize all curricular development.

Regarding the need for dissemination, attendees suggested the center could serve as a hub for reviewing and sharing curricula and educational models.  For example, building on prior successful models (e.g., PhysPort\cite{physport}, nanoHUB\cite{nanohub}) a center could host a repository of quality and vetted educational resources for educators and the public alike. 

Similar to curriculum development, a center would also be well-positioned to create, host, and share informal activities for educators and students, as well as informational materials for different audiences. At the K-12 level, there is some overlap between informal activities and curricula. This is still a new area with many projects focused on games and science kits as a fun, quick way for educators to include quantum concepts in their classrooms. However, the current offerings are relatively narrow in scope and many are  insufficient for immediate adaptation into curricula. A center could serve to assist, when applicable, in the development of formal scaffolding for curricula. A center could also shed light on the need for a more expansive set of informal activities and experiences that cover more QISE topics, and possibly develop partnerships with museum educators or other experts in this domain. Due to their applicability to formal classrooms, the informal resources should be curated in the curricular repository described above (maintained by a center), and be distributed throughout a network, such as the National Q-12 Education Partnership's QuanTime program \cite{Q12}, which collects and disseminates educational materials and guides to teachers for programs such as World Quantum Day.

Attendees also noted that projects aimed to have broader impact are a mandatory component of many research centers, especially those funded by the NSF. However, there were questions regarding the quality and efficacy of the public engagement activities that are designed by research-first programs without educational experts. While not necessarily a central feature, a quantum education center could serve as an additional resource for researchers to leverage such that they could avoid starting from scratch, contribute novel ideas where there are gaps, and improve their educational programs. On the other side, research programs with robust, high-quality efforts that broaden participation could leverage the center to amplify their reach with a wider audience. This two-way exchange could better ensure that quantum educational programs and institutions sustain their connections with students from diverse geographies, identities, and socioeconomic backgrounds.

\textbf{Evaluation and Assessment} Participants recognized that evaluation plays a key role in the success of educational initiatives. This center is no exception, and could include evaluation at multiple scales: from developing and curating resources for evaluation of individual programs, to assessing and evaluating impacts of center-supported initiatives, to analyses of the landscape of quantum education nationally. Within individual or small-team programs, the development of validated assessments to measure impacts is costly and time-consuming. A center could help alleviate this by supporting, curating, sharing, and in some cases, developing, appropriate assessments to advance initiatives aimed at quantum education. At the same time, common approaches to evaluation of education will allow coordination and comparisons across currently isolated efforts.  

In parallel, a center should commit to conducting validated forms of assessment to understand what is happening and how to improve activities, and evaluation to document the impacts and outcomes of its initiatives. A center's professional development program, for example, would likely benefit from a large-scale assessment across regional cohorts; a center has access to more participants, which would mean more data, allowing for the overall refinement of programs and documentation of impact.

A secondary pillar of assessment relates to the landscape of quantum education. A center would be well-positioned to periodically collect and analyze data related to the workforce, such as numbers of students engaged in quantum programs, changes to public perception and awareness, and the needs of future employers. By monitoring workforce attributes and needs of educators, students, and employers, the larger QISE community will be able to better adapt educational programs and resources accordingly.

As part of this discussion, participants began to discuss possible metrics of success. This is challenging due to the hypothetical nature of the proposed educational investment, but the following examples were identified as possible metrics:
\begin{itemize}
    \item increase in number of students expressing interest in quantum-related careers 
    \item increase in educators participating in QISE-related PD
    \item increase in number of states including quantum concepts into STEM courses as measured by changes to standards or publicly available curricula (K-12)
    \item increase in number of quantum programs reporting sustained K-16 student participation
    \item increase in publicly available curricular resources across all age groups K-16
    \item increase in the number of QISE courses offered at 2- and 4-year colleges
\end{itemize}

\textbf{Programs for scaling accessibility to hands-on and/or laboratory equipment} The discussion highlighted that careers in QISE often demand a strong foundation in experimental knowledge and hands-on skills. However, the high cost and complexity of the necessary laboratory equipment often makes these educational opportunities inaccessible to many students. To address these challenges, a center could play a pivotal role in reducing barriers and enhancing the integration of experimental education into the undergraduate curriculum.

Firstly, the center could compile and disseminate information on cost-effective alternatives that still meet the desired learning objectives. By identifying and promoting affordable yet effective equipment and experimental setups, the center can help institutions overcome financial constraints while maintaining the quality of education.

Secondly, the center could facilitate collaborative purchasing among a consortium of institutions. By organizing group purchases, institutions can benefit from economies of scale, reducing the cost per unit of equipment. This approach would enable more widespread access to the necessary tools for experimental learning.

Additionally, the center could act as a bridge between academia and industry, fostering partnerships that make remote access to advanced equipment more feasible. By leveraging industry resources, the center could enable students to engage with cutting-edge technology without the need for direct ownership or on-site presence. This could include remote laboratory setups or shared-use facilities, all of which would provide students with essential experimental experience.

At the K-12 level, a center could play a similar role in identifying age-appropriate labs or hands-on activities for each grade-band and core STEM subject. It could then curate cost-effective options matching the learning goals and expectations. Group purchasing would be more challenging for pre-college institutions, but the center's curated options would provide a starting point for states, districts, and schools with an interest in including hands-on quantum modules (e.g., labs, experiments).

\textbf{Network building}   While there is currently sharing across the quantum education community, the attendees noted that building capacities to connect people, programs, and models for advancing QISE education serve as essential forms of infrastructure that a national center could sustain. Whether program leaders, practitioners, industrial members, students, or interested stakeholders, participants from these varied communities all benefit from discussions within and across these sector boundaries. Intentional alignment and coordination across the programs in QISE will provide enhanced pathways and opportunities for the wide array of learners impacted by and impacting the future of quantum education, while simultaneously making programs more robust, resilient, scalable, and responsive to local and national needs. Additionally, a national center has the unique capacity to curate and coordinate knowledge about how various models work, where and when, and what gaps are needed to be filled. Proactively developing and sharing knowledge about effective approaches ensure enhanced opportunities and adaptable, sustainable plans for a vibrant quantum ecosystem for decades to come. 

\textbf{Public Engagement and Science Communication} 
Attendees noted that establishing a center in quantum education is necessary to elevate the visibility of the challenges and possible solutions in the domain of broad quantum workforce development. This visibility is necessary to bring collaborators and decision makers into the community in a way that positively affects change at scale. For example, in the long-term, quantum education should be well integrated into STEM education at all levels-- this is not possible without community engagement to drive meaningful collaboration and connections to existing STEM infrastructure. A center also provides a clear mechanism for experts in quantum education to be on equal footing with leaders of large-scale research centers in the domestic and international conversations around the future of quantum science research and technology. 

Beyond visibility, the attendees discussed the persistent lack of public information around QISE and the associated careers and applications. While there are currently small programs that focus on advancing quantum literacy among members of the public, attendees cited a need for a more coordinated informational campaign or set of resources tailored to different audiences. It is feasible that a center could facilitate a high impact effort in this area, and reach a diverse audience of students, educators, and families. For example, a broader public engagement presence could support increased awareness and interest among parents and others who support students.

A center focused on education would, due to its inclusion of K-12 and early undergraduate, likely develop wealth of expertise in approaches to discussing QISE topics to non-experts. Therefore, it could also act as a resource to share best practices of science communication in QISE to help educators, professional scientists and engineers, and research centers communicate about this critical area of STEM to a larger, more diverse audience.

\section{Outcomes from establishing a quantum education center}

The primary outcomes of a center-scale investment in quantum education would include (1) an increase in the number of educators prepared to teach quantum-related topics; (2) an increase in the number and diversity of students aware of opportunities in, and prepared to enter, the quantum workforce; (3) strengthened pathways into quantum careers; (4) a more complete understanding of the demographics and distribution of the population prepared to enter the US quantum workforce as it changes in the coming decade; and (5) increased public awareness and engagement with quantum information science, engineering, and technology. 

Collectively, these outcomes will have a significant impact on the state of quantum workforce development. A national center that includes the activities listed in the section above would be well positioned to deliver the data, programs, and support the community needs to prepare students across different regions for jobs in quantum science. It would also provide a sustained mechanism for tracking national participation, monitoring needs, and seeding programs that address gaps. By comparison, isolated regional efforts are presented with a significant challenge to tracking effectiveness over long timescales, especially when the programs take place at earlier stages; a national-scale center will not only be able to scaffold these educational efforts, but report on outcomes nationwide. 

 A center that facilitates connections across programs will also enhance the QISE community, helping to build a wider range of educational pathways into quantum careers beyond what is currently available. For example, regional efforts in QISE education would be better able to connect with each other and other national centers to amplify their needs, impacts, and opportunities across the country and globe. A center also would provide a more efficient mechanism to connect education efforts to industry needs, helping to create a robust feedback loop of employers and educators that ensures students are prepared for whatever lies ahead in QISE.

Critically, the attendees felt that a center would provide the necessary infrastructure to ensure that there is collaboration across the activities in Section 4.2. This integrated approach is necessary to achieve the broad outcomes listed above. For example, combining PD, curricular development, evaluation, access to equipment, a connected network of educators, and public engagement will result in a expansive community of practice that supports educators across more classrooms and education sites nationwide. Over the long term, an integrated approach will lead to increased awareness, confidence, and knowledge of QISE among more educators across multiple geographies and disciplines.

In addition, supporting the creation and dissemination of educational practices in formal and informal quantum education across the community may lead to better outcomes for broader impacts activities in QISE research programs beyond a center. Many attendees noted that while ensuring the quality of broadening participation programs is not specifically a QISE issue, this center could have significant impact through curation and dissemination activities. For example, QISE draws on fields with some of the lowest percentages of both women and people who are African-Americans, Black, and/or Latine \cite{physicsraceethnicity2018, NCSESstats}. By sharing information about programs that are successfully removing barriers to QISE for these groups, a center could accelerate the wider adoption of such practices while the field is relatively young, rather than forcing a difficult pivot when QISE is more established.

In summary, workshop attendees widely agreed that establishing a national center is a next step that could accelerate the widespread adoption of QISE education. Specifically, such a center could operate across the ecosystem to support cohesion, collaboration, and implementation at scales not possible with individual investigators or institutions. Ultimately, such work could help inform future investments in areas of QISE workforce development, and provide additional guidance on how to leverage the vast educational and research infrastructure across the nation to prepare an agile, quantum-ready workforce.

\section{List of abbreviations}
\begin{itemize}
    \item QISE: Quantum Information Science and Engineering
    \item HBCUs: Historically Black Colleges and Universities
    \item NQIAC: National Quantum Initiative Advisory Committee
   \item NQIRA: National Quantum Initiative Reauthorization Act
   \item NQIA: National Quantum Initiative Act
   \item PD: Professional Development
   \item NSF: National Science Foundation
\end{itemize}

\backmatter

\section*{Declarations}

\begin{itemize}
\item Funding
The workshop was supported by the National Science Foundation Grant numbers OSI-2405381 and PHY-2317149.
\item Competing interests:
E. Edwards provided unpaid expert testimony related to quantum education to the US House of Representatives on June 6, 2023, on the request of members of congress. Edwards is an unpaid member of the state of Colorado's Elevate Quantum Workforce Advisory Board.

B. Brown is Executive Director of QuSTEAM, a nonprofit organization emerging from an NSF Convergence Accelerator project. QuSTEAM offers a platform for QISE educators and researchers to collaborate on curriculum development and has economy-of-scale contracts with all willing providers of physical and remotely accessed quantum devices.

\item Ethics approval and consent to participate:
Not applicable
\item Consent for publication:
All authors gave written permission to be included as an author on the manuscript. 
\item Data availability: 
Not applicable
\item Materials availability:
Not applicable
\item Code availability: 
Not applicable
\item Author contribution:
E. Easton, E. E. Edwards, N. Finkelstein, and H. Lewandowski drafted and edited the manuscript based on detailed workshop discussion notes and presentations. All authors participated in either the in-person workshop or virtual meetings, as well as gave feedback on drafts of the manuscript. The workshop organizers acknowledge and thank all the attendees of the workshop for their participation and for the valuable discussions that resulted in this report.
\end{itemize}

\noindent


\end{document}